# Characterizing Algal blooms in a shallow and a deep channel over a decade (2008-2018)

Maryam R. Al Shehhi[1]*, David Nelson[2], Rashid R Alkhori, Rashid Alshihi[3], and Kourosh Salehi-Ashtiani[2]

*1 Civil, Infrastructure, and Environmental Engineering Department, Khalifa University, UAE, Abu Dhabi*
*2 Laboratory of Algal, Systems, and Synthetic Biology, Division of Science and Math Center for Genomics and Systems Biology, New York University Abu Dhabi*
*3 Ministry of Environment and Climate Change, UAE*

**Abstract**

The outbreaks of algal blooms occur in both shallow and deep-water bodies. To compare the characteristics of algal blooms in the shallow and deep water, we consider the Arabian Gulf and Sea of Oman as a case study. While the Arabian Gulf is a shallow region dominated by advective features, the Sea of Oman is a deep channel dominated by numerous eddies. Both these regions is have rarely been studied due to lack of available data thus preventing the characterization of phenomenon such as algal blooms in the regions. Nonetheless, a recent unique and comprehensive dataset of the frequent algal blooms collected over the last decade in the Arabian Gulf and Oman Sea is utilized in this study to analyze the spatio-temporal variability of the blooms thereof. These data are also used to characterize the algal bloom species and analyze the relationship between the water properties and the algal blooms in the shallow and deep waters. There is a general decreasing trend of the algal bloom events from 2010 to 2018 in the Arabian Gulf while in the Sea of Oman there is an increasing trend. We reveal a clear seasonality with the highest frequency of algal blooms during winter and spring. We have noticed that algal blooms have the feature of the annual cycle with initial blooms happening in November-December and December-January in the Arabian Gulf and Sea of Oman, respectively. The analysis further demonstrates that the algal blooms grow better at salinity levels of 39-40 psu/37-37.5 psu, temperature of 23-24 °C, and pH of 8 in the Arabian Gulf/Oman Sea. Findings of this study provide insight into the relationship between water properties and algal bloom frequency, and a basis for future research into the drivers behind these observed spatio-temporal trends.

1. **Introduction**

Algal blooms have been observed frequently in the shallow and deep oceans worldwide (Sellner et al., 2003). While these microscopic organisms are significant source of food and oxygen to the aquatic system, their over-growth can cause adverse effects to the aquatic system (Al Shehhi et al., 2014). Some of these species can produce toxins that accumulate in marine organisms (e.g. fish) and consequently cause illness to humans. These illnesses include: 1) Amnesic Shellfish Poisoning (ASP), 2) Paralytic Shellfish Poisoning (PSP), 3) Ciguatera Fish Poisoning (CFP), and 4) Diarrhetic Shellfish Poisoning (DSP) (Anderson, 2009; Glibert et al., 2005) which are caused by the toxins Domoic acid, Saxitoxins, Ciguatoxin/Maitotoxin, and Okadaic Acid (Van Dolah, 2000) respectively. Marine inhabitants could also be killed by the extensive consumption of oxygen by the algal species. The algal species are also responsible for degrading the quality of the water by causing water discoloration, and disturbance of industrial activities such as the desalination plants by clogging their filtration and salt removal systems (Al-Shehhi et al., 2014; Richlen et al., 2010).

These drawbacks of the algal blooms are more serious in the shallow coastal waters than the deep waters due to a generally high intensity of blooms in the shallow waters and its negative impact on the coastal communities (Boesch, 1997)(Richlen et al., 2010). This is explained by the availability of the sunlight and nutrients in the shallow regions that support the algal growth as found in previous reports. Most of these studies were focused on the shallow and deep lakes, not the open oceans. The lakes and open-oceans are different in terms of the water dynamics and movement, which significantly affects the ecology of algal blooms(Clasen et al., 2008). Therefore, in this study, we aim to characterize the algal blooms in the shallow and deep open-oceans. For this purpose, we consider the Arabian Gulf, which is a shallow region dominated by advective features, and the Sea of Oman which is a deep channel dominated by numerous eddies as a case study (Reynolds, 1993). In addition to the interesting oceanic and topographic features of these regions, numerous algal bloom events have occurred in this region in the last three decades (Al-Ansi et al., 2002; Glibert et al., 2002; Heil et al., 2001; Zhao et al., 2015). Most of these bloom events were caused by dinoflagellates (Al-Shehhi et al., 2014). These algal bloom events have caused several drawbacks. For example, over 7,000-10,000, 2,500, 100, and 650 tons of fish were killed in Salalah (Oman), Kuwait, Duqm (Oman), Fujairah (UAE: United Arab Emirates) due to the depleted- oxygen waters in 1976, 2001, 2004, and 2008, respectively (Al-Gheilani and Matsuoka, 2011; Berktay, 2011; Heil et al., 2001; Piontkovski, 2012). Waters colors turned to red, green-orange, yellow-greenish, and brown in 1999, 2008, 2013, 2016 due to the presence of *Mesodinium rubrum* (Al-Busaidi et al., 2008; Heil et al., 2001), *Trichodesmium*, *Prorocentrum micans*, diatoms (*not specified*) and *Noctiluca respectively*. Moreover, three membrane-based desalination plants were sealed off in Ras Al Khaimah and Fujairah in 2008-2009, while Kalbah desalination plant was sealed off in 2013 due to clogging the membrane modules (pore size of 0.0001 – 0.001 µm) (Membrane Research Environment (MemRE), 2017) by the algal species whose diameters vary from few to hundreds of microns(Rogers, 2011).

The growth of these algal species in this region is found to be more frequent in the spring season due to the availability of nutrients. The sources of these nutrients could be either natural (e.g. seabed nutrients, and deposition of atmospheric dust), or artificial (e.g. waste of coastal human activities). The seafloor nutrients, such as nitrogen/Ammonium and phosphorus (Gruber, 2008; Kraal et al., 2012; Murty et al., 1968), could be brought up to the surface by winter convection and water circulation and create a good environment for the algae at the sea surface. In addition, the frequent dust storms, usually caused by the *Shamal* wind, a northwesterly summer wind in the Arabian Gulf, transport heavy dust particles, which can get deposited into the water surface. These dust particles contain mainly calcium carbonate, and heavy metals such as aluminum and reactive iron (Hamza et al., 2011; Lyles et al., 2008; Thalib and Al-Taiar, 2012). Apart from these natural nutrients, the coastal islands' reclamations activities and the coastal industrial plants provide the industrial nutrients(Al-Shehhi et al., 2014). For instance, more than 30 wastewater tankers were caught to discharge the sewage water (bio-waste) illegally into the gulf (Al-Shehhi et al., 2014) while highly contaminated waters were observed in the industrial southern region of the gulf (Abu Dhabi: Mussafah, UAE).

Due to all the aforementioned serious impacts of the previous algal bloom outbreaks in the region, the regional environmental entities had established plans to start collecting the data that is important to monitor the blooming activities along its shores and reduce the risk of future blooming events. The United Arab Emirates (UAE) is one of the countries that devoted a lot of efforts to build high temporal sampling system to monitor the water quality along its shores and reduce threat to the public health. The main environmental entities, which are in charge of monitoring the marine water in the UAE, are the Ministry of Climate Change and Environment (MOCCAE), and the Environment Agency of Abu Dhabi (EAD). The monitoring plans of MOCCAE and EAD are focused mainly on the northern and southern coasts of the UAE, respectively. These environmental entities have recorded all the algal bloom events that have occurred and the corresponding water properties (e.g. temperature, salinity, dissolved oxygen, and pH) since 2008.

The unique dataset collected has provided the opportunity to characterize the algal bloom activities along the shallow coasts of the UAE, which lie on the Arabian Gulf and the deep waters of Sea of Oman, over a decade (2008-2018). The outcome of this study includes and are not limited to:

1. Studying the inter-annual and seasonal variability of algal blooms over a decade.
2. Characterizing the initial algal blooming events.
3. Estimating the toxicity levels and chemical compounds of these species.
4. Studying the relation and influence of the water properties on the algal productivity in this region.

2. **Materials and Methods**

2.1 Site Characteristics

The Arabian Gulf is a semi-enclosed basin located between 24º N and 30º N and longitude 48º E and 57º E (Figure 1). Its eastern part is open to the Sea of Oman, which joins the Arabian Sea and the Indian Ocean. While the Arabian Gulf is mostly shallow with an average depth of 30 m, the depth of the Sea of Oman reaches up to 3000 m (Pous et al., 2004). The area surrounding the Arabian Gulf is hot and dry most of the year (March-October) where the sea surface temperature can reach 35 ºC in the summer and the precipitation rate is very low (below 300 mmyr$^1$) (Alsharhan et al., 2001). Additionally, this region has frequent dust storms (15-20 per year). These dust storms are usually caused by the winds blowing from the Arabian Sea (Shamal wind), Northeastern Africa and Afghanistan (Gherboudj and Ghedira, 2014; Jish Prakash et al., 2015). Due to the occurrence of these winds and the water density gradient, a clockwise circulation occurs in the middle of the Arabian Gulf, while the water flows mainly from the Sea of Oman into the Arabian Gulf and continues upward until it reaches Kuwait, then changes its direction downward until it reaches Sea of Oman again (Bjerkeng and Molvær, 2000).

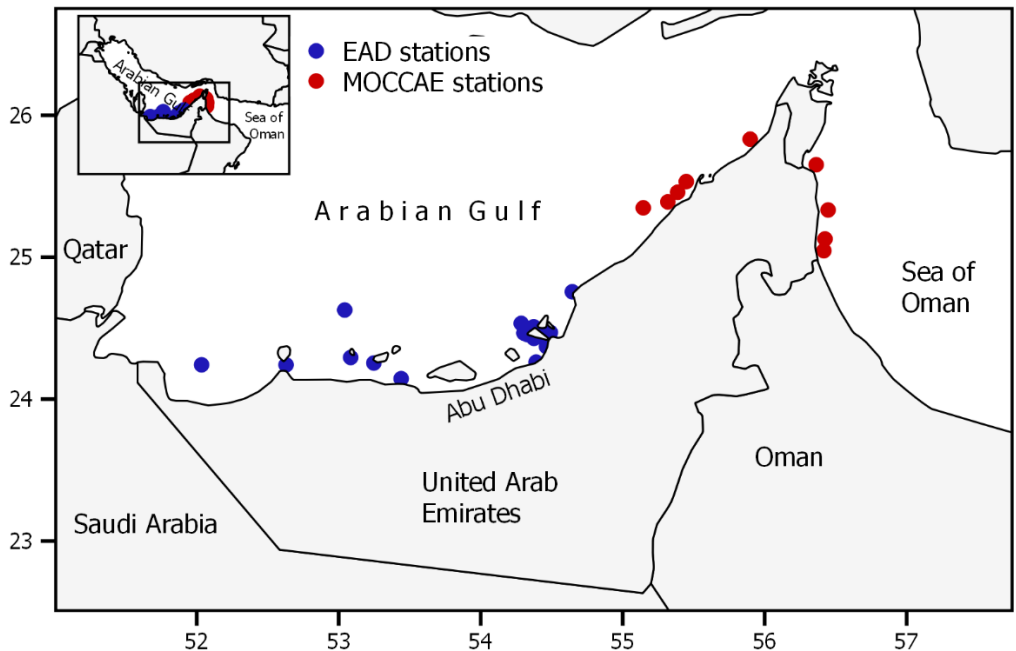

Figure 1. Map of the Arabian Gulf and Sea of Oman, and the location of the in situ stations along UAE coasts.

The combination of these unique features (exclusive geographical structure, harsh meteorological conditions, and frequent wind) cause low water renewal rate and high evaporation rate of the gulf water, which consequently cause intricate chemical properties such as the high salinity which has an average and maximum values of 43 psu and 80 psu, respectively. This extreme in salinity, as well as in temperature, affects the biological properties of the region (Naser, 2014; Riegl and Samuel, 2011).

2.2 Data processing

*MOCCAE Dataset and samples*

All the algal blooms events occurred since 2008 have been recorded by MOCCAE at 9 stations (RAK: Ras Al-Khaimah, UAQ: Umm Al Quwain, Ajm: Ajman, Shj: Sharjah, DUB: Dubai, Dibba, KF:Khor Fakkhan, Fujairah and Kalba) shown in Figure 1. The coincident water quality properties: temperature (ºC), dissolved oxygen (mg L$^{-1}$), salinity (psu), and pH have been recorded as well. This data has been used to analyze the interannual variability of algal blooms along the northern coast of UAE, as well as, the seasonal and spatial variability. The analyses include temporal linear trend fitting and two-pass filtering. The geospatial platform (QGIS) has also been used to map the algal bloom events over four seasons (DJF: December-January-February; MAM: March-April-May; JJA: Jun-July-August; SON: September-October- November).

In addition to the abovementioned dataset, samples of some initial algal bloom events occurred in 2015-2016 have also been used to perform toxicity and morphological analyses. For the toxicity analyses, High Performance Liquid Chromatography (HPLC) was used to measure Chlorophyll a (Chl-a) concentration, primary metabolisms and biologically active compounds. The Autosampler used was set to draw 2 µL per run. A C18 column was used with a flow rate of 0.300 mL min$^{-1}$. We created a custom solvent gradient based on of the work of (Castro-Perez et al., 2010) utilizing a starting solution of 28% acetonitrile, 36% water with 30mM ammonium formate, and 36% isopropanol (IPA). This was changed as a gradient to 80% IPA and 20% water over 40 minutes followed by 2 minutes of flushing with the starting solution. The HPLC-MS was operated in positive multiple ion detection (MID) mode with two set reference lock mass compounds for accurate compound profiling. However, for the morphological analyses, we have obtained 2-D and 3-D images of the algal species by using the light and the scanning electronic microscope (SEM). To achieve high quality microscopic images, 200-1000 mL of the seawater were filtered first using 0.45 µm (pore size) GF/F filters under vacuum and then fixed with Lugol (20%) for 24 hrs. Two additional steps were considered for the SEM, washing the samples three times with distillated water and drying it by a freeze-drier.

*EAD Dataset*

Monthly water quality data for 22 stations were obtained from EAD along the southern coast of the UAE, Abu Dhabi, for the period 2010-2011 as shown in Figure 1. The collected water quality data includes: chl-*a* (mg m$^{-3}$), temperature (°C) and dissolved oxygen (mg L$^{-1}$). Chl-*a* is the indicator of the occurrence of algal bloom events as the algal blooms increase the aquatic content of Chl-*a*. This dataset was used to analyze the monthly trend of algal blooms in the southern coast of UAE. This data has been also used to study the dependency of the blooms on the water properties (temperature and dissolved oxygen). The raw data has been clustered based on the various levels of Chl-*a* associated with temperature and dissolved oxygen. Figure 2 shows the flow chart of the methodology being followed in this work.

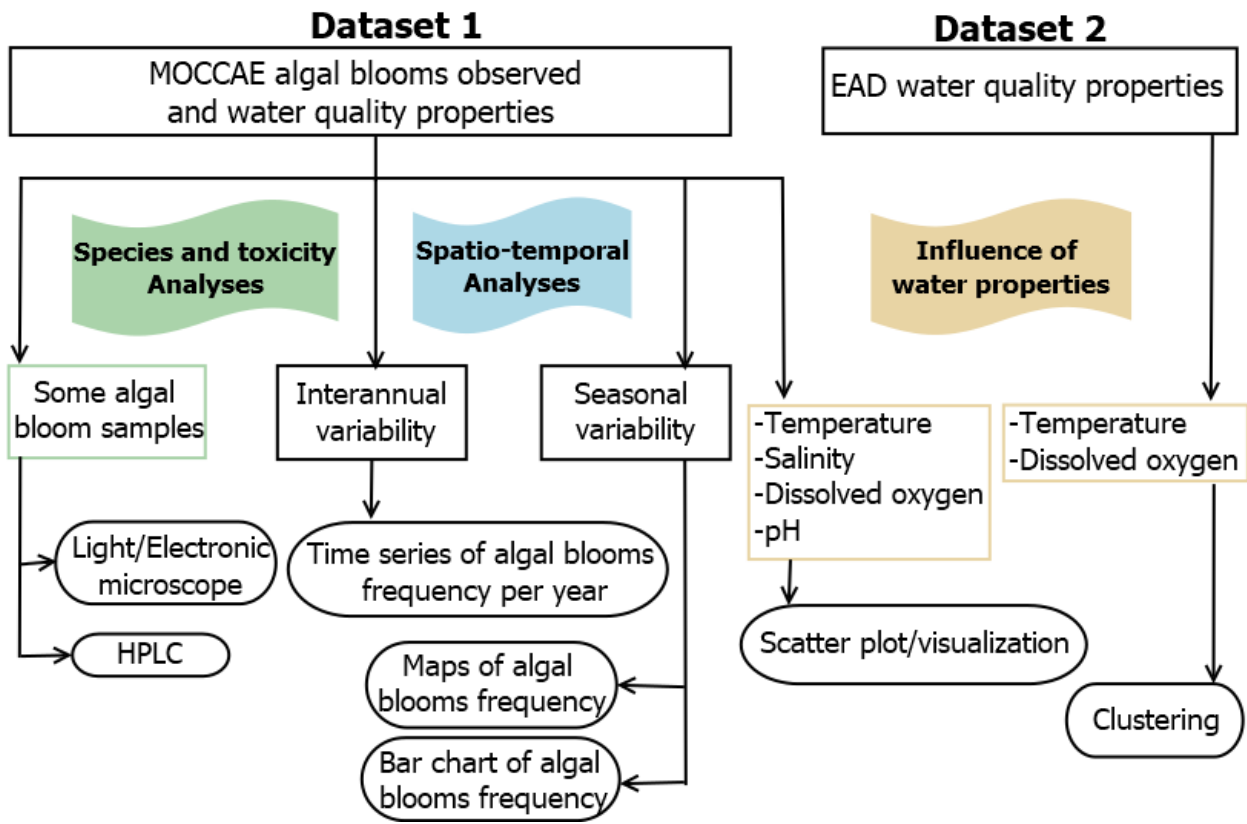

Figure 2. Methodological flowchart of algal bloom analyses along the coasts of UAE.

## 3. Results and Discussions

3.1 Inter-Annual variability of algal blooms

Figure 3 presents a moving average (two-pass) filter applied to the total number of algal blooms recorded along the western and eastern coast of the UAE between the years 2010 and 2018. This figure highlights the temporal trend of the annual frequency of algal blooms as recorded along the coasts of the UAE by using the linear regression line of the applied two-pass filter. As shown in this figure, the highest frequency of algal blooms occurred in the eastern coast of the UAE are those of 2014 based on both the two-pass filtered and unfiltered data of the algal bloom frequency. However, in the western region, the highest annual total numbers of algal bloom events were recorded in 2012 and 2011 based on the filtered and unfiltered data, respectively. In contrast, Figure 3 highlights that the lowest number of algal bloom events were recorded in the years 2016, 2017, and 2012, 2018 in the western and eastern coast of the UAE.

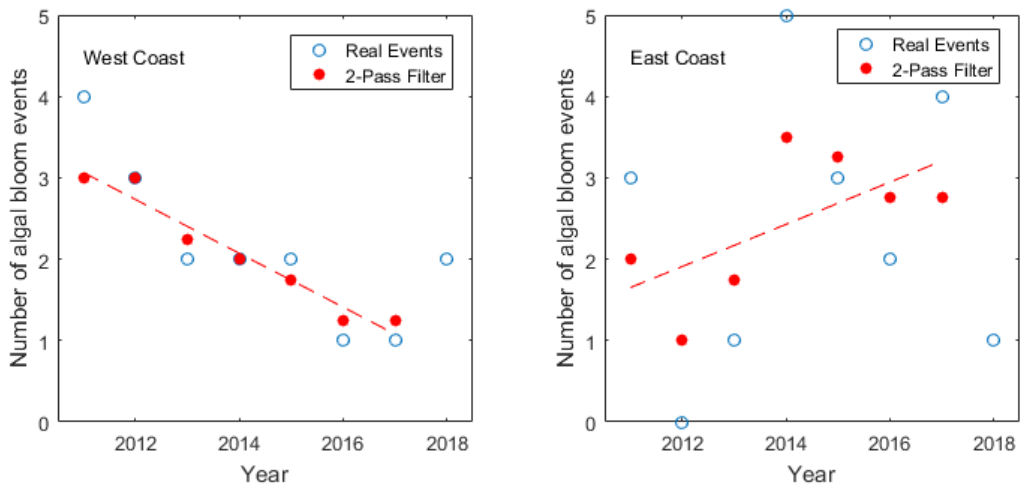

Figure 3. Temporal trend analysis of the total annual number of algal bloom events as recorded over all the in situ stations (MOCCAE) along the east and west coasts of UAE over the years 2010 and 2018, including a 2-pass filter and linear regression line.

From Figure 3, it is also evident that the variation in the annual frequency of algal bloom events varies in space as well. For example, in the western coast of UAE, there is a general decreasing trend of the algal bloom events from 2010 to 2018 based on the filtered data fit. It dropped from 3 events per year in 2010 to less than 2 events per year in 2018. In spite of this decreasing trend, in 2018, algal bloom events occurred twice. On the other hand, there is an increasing trend of algal blooms events in the eastern region over the same period (2010-2018). It varied from 3 to 5 algal bloom events over the 10 years.

This fluctuating annual trend of the algal bloom events' frequencies in the eastern-western coasts of the UAE is controlled mainly by the geolocations of these coasts. More frequent algal blooms occur in the eastern coast (Sea of Oman) due to the high levels of nutrients (Nitrate >0.4µm;Phosphate >0.4 µm), compared to the western coast (Arabian Gulf) of lower level of nutrients (Nitrate <0.5µm;Phosphate <0.4 µm). These levels of nutrients have seasonal variability as well in the Arabia Gulf and Sea of Oman due to variability of deposited iron, from seasonal dust storms, and the upwelled nutrients, affecting the seasonality of the algal blooms.

3.2 Spatio-Seasonal variability of algal blooms

Considering the 3d bar-charts (Figure 4) of the monthly frequency of algal blooms between the years 2010 and 2018, the most noticeable feature is that the longer period of high-frequency algal blooms in the winter and spring seasons between the months of November and April. In addition, we have noticed a new feature of the blooms in this region, which is called the initial blooming. Initial blooms refer to the first occurrence of blooms in the annual cycle. It is interesting to note that initial blooms used to occur in November in the west coast, but it got delayed in the years between 2016 and 2018 to start in January. In contrast, the initial blooming occurred in December in the east coast over the whole decade.

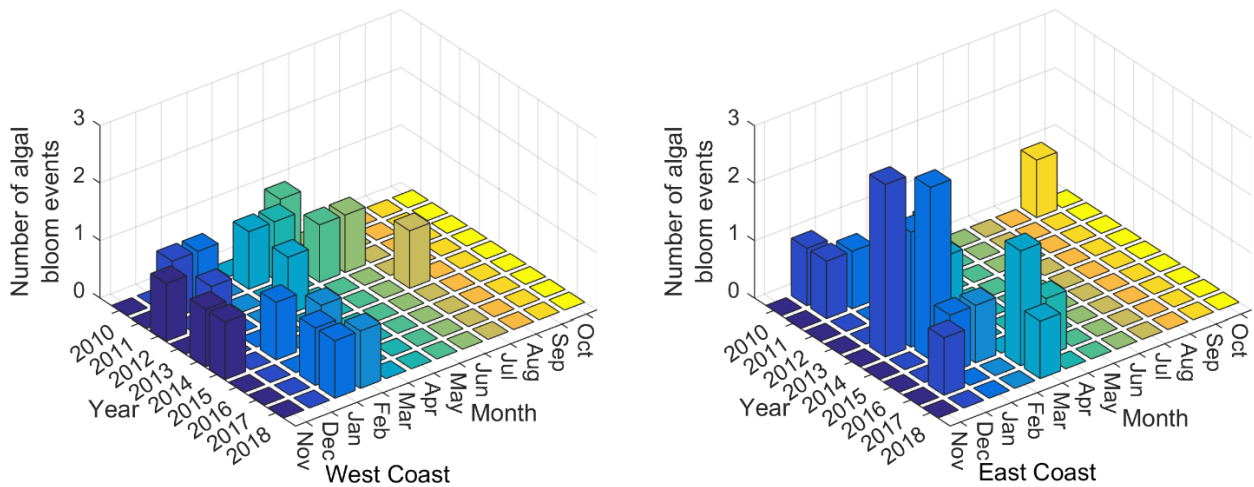

Figure 4. Monthly frequency of algal blooms as recorded over all the in situ stations (MOCCAE) along the east and west coasts of UAE over the years 2010 and 2018.

It is evident that, the blooms were intensive and occurred for a longer period of time (November-May) in the shallow west coast for the year 2010-2013. However, after 2014, the blooms occurred for a shorter period of time (December-March). In the deep eastern coast, blooms maintained their appearance between December and April almost over the whole decade. While the highest frequencies of blooms in the east coast have reached up to 2 events/month in December and January whereas in the west coast the highest frequency of blooms was 1 event/month.

In addition to the high occurrence of the algal bloom events during the winter and spring seasons, the events have shown a spatial variability within these seasons. Figure 5 shows the distribution of algal bloom frequencies over four seasonal periods (DJF: December-January-February; MAM: March-April-May; JJA: Jun-July-August; SON: September-October- November). The blooms are commonly recorded at nine stations in which five stations are located in the west coast of UAE, and the rest are in the east coast. These nine stations are labeled in Figure 5 corresponding to their geolocations. As shown in this figure, during the DJF, the most noticeable feature is that algal bloom frequency reaches its maximum peak in *Dibba* with total number of seven algal bloom events. During the same season, algal blooms occur in *Kalba* and (UAQ: *Umm Al Quwain*) as well more than three times at each station. While the frequency of blooms is high in the DJF season, they have shown less spatial distribution in this season compared to the MAM season. During the MAM season, blooms occurred at all the nine stations with maximum algal bloom events of five per station recorded in the east coast. It is also very interesting to see that during the SON season, blooms mainly occur in the west coast and not being influenced or transported from the Sea of Oman due to the Indian southwest monsoon that activates a lot of upwelling areas in the Sea of Oman and Arabian Sea.

3.3 Characteristics of the initial blooming events

As mentioned in section 3.2, the initial blooming occurs in November-December and December-January in the west and east coasts of the UAE, respectively. To characterize the initial blooming in this region, we have analyzed the initial blooming event that occurred in November in the west coast of UAE. We have collected algal bloom samples during the event of November 2015. The blooms were initially observed in November 2015 in the western coast of Ajman with high Chl-*a* exceeding 2 mg.m$^{-3}$. However, downward water currents with speed of 9-12.4 cm/s transported the blooms into the southern region toward Sharjah, and Dubai. Diatoms mainly caused the bloom. These species were responsible for discoloring the seawater to yellowish-greenish color.

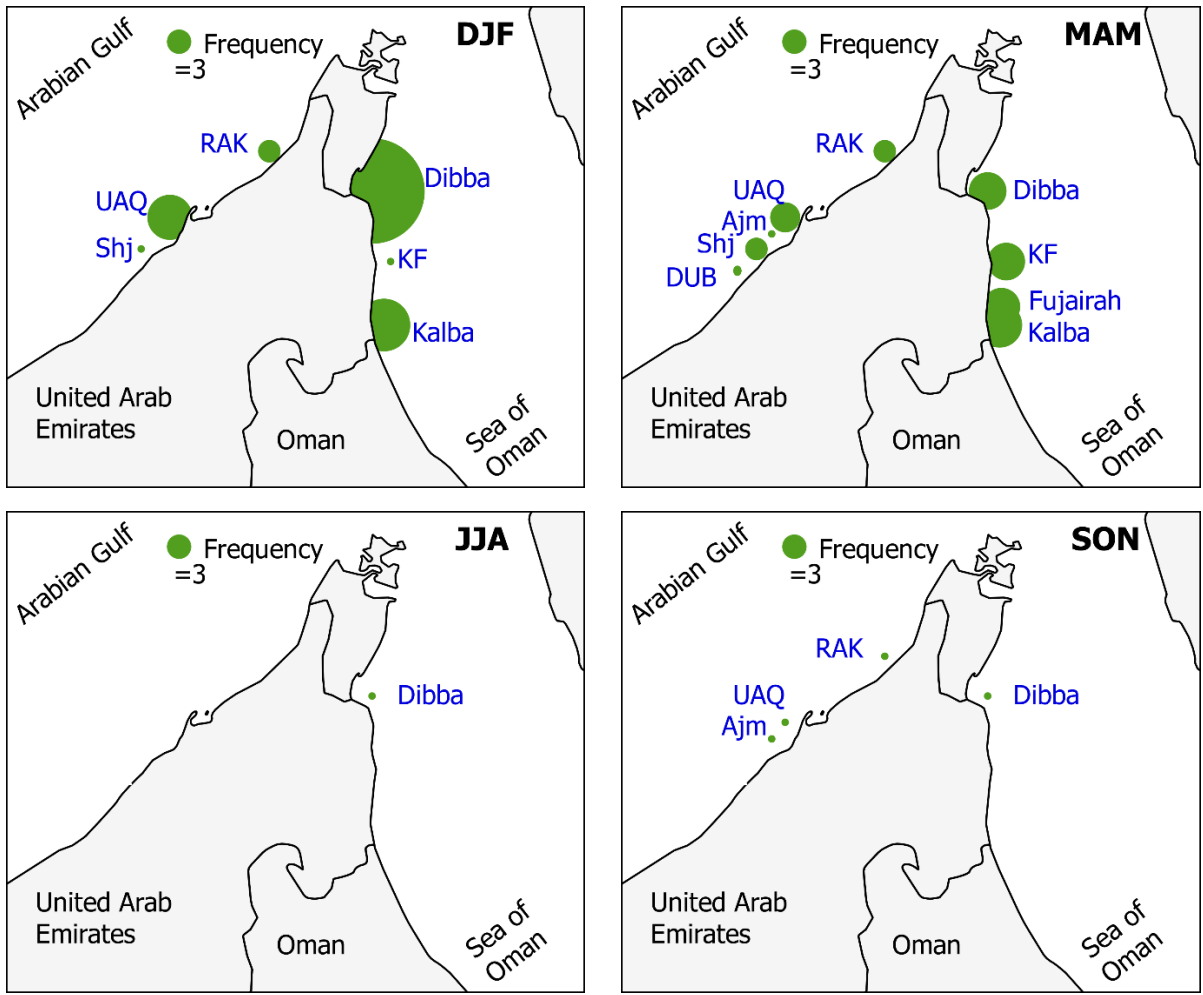

Figure5. Maps of seasonal frequency of algal blooms as recorded over all the in situ stations (MOCCAE) along the coasts of UAE over the years 2010 and 2018: (DJF: December-January- February), (MAM: March- April- May), (JJA: Jun-July-August), and (SON: September-October-November).

In addition, a dinoflagellate bloom event that occurred in UAQ in February 2016 will be analyzed in this section to compare the chemical composition of diatom and dinoflagellate bloom events. Here we identify the initial blooming species and analyze their toxicity and chemical compounds.

*Morphological Identification*

The initial algal blooms that occurred in November 2015 have been caused mainly by diatomic species. We were able to identify four diatomic algal species dominating the water samples which are: *Thalassionema,*

*Chaetoceros, Thalassiosira*, and *Pleurosigma* during the initial blooming events. All these species have been identified by the light and electronic microscopy. Figure 6 shows several microscopic images of these four species. *Thalassiosira* species made up 27%, and 24% of the algal bloom samples in Ajman and Dubai. These species prefer living in the warm waters. For example, higher concentrations of *Thalassiosira* species were in Omani waters in 2010 and 2011 during the end of southwest monsoon (October) when sea temperature reached 30 °C (Al-hashmi et al., 2014). In addition, *Thalassionema nitzschioides* and *Thalassionema frauenfeldii* were numerously found in the seawater samples collected from Ajman (36%) and Dubai (28%). These species prefer living in cold seawaters at an average temperature of 15 °C. They had previously caused less concentrated algal bloom events in the Arabian Gulf in October 1999 (Kuwait waters) at sea temperature range of 26.9 °C and -28.6 °C (Heil et al., 2001), and higher concentration algal blooms in Omani waters in November 2008 at lower sea temperature range of 23°C and 24 °C (Piontkovski et al., 2011).

*Chaetoceros* species were found numerously in the seawater samples collected from Sharjah (34%). These species live in the warm waters and they were found to grow best at 30 °C (Adenan et al., 2013), and its abundance has increased from $5.47 \times 10^3$ cell.L$^{-1}$ to $24.84 \times 10^3$ cell.L$^{-1}$ due to an increase of sea temperature from 20 °C to 25 °C at the coastal waters of Pakistan (Khokhar et al., 2016).

On the other hand, the dinoflagellate (*Noctiluca scintillans:*no picture) were found with extensive concentration in UAQ in February 2016 at 21 °C. These species commonly cause more than 50% of the blooms in the Arabian Gulf (Al-Hashmi et al., 2015) because their best growth rate is at sea temperature between 24 to 27 °C (Al-hashmi et al., 2014). These species also have the ability to produce ammonia ($NH_4^+$) that is possibly toxic to the fish, thus, an extreme $NH_4^+$ was detected in this location (551 mg L$^{-1}$).

*Toxicity analysis*

Toxic biological compounds (Chl-*a* and accessory pigments) of the initial blooming species are analyzed in this section. This analysis is undertaken by HPLC-MS. Based on the HPLC-MS analysis, it is determined that samples collected in Ajman and Dubai contain high concentrations of phytoplankton bloom metabolites as shown in Figure 7a. It is verified by detection of Chl-*a* and Chl-*b* with [M+H]$^+$ ions (m/z: 893.54) and [M+CH3OH+H]$^+$ ions (m/z: 906.5), respectively. The corresponding Chl-*a* values are 4.85 ± 0.56 mg.m$^{-3}$, and 3.77 ±1.59 mg.m$^{-3}$, which could reflect the peaks obtained by the HPLC-MS. These chlorophylls are the results of photosynthesis processes in the diatomic species *Thalassionema*, and *Pleurosigma* found in Ajman and Dubai waters. Ajman seawater showed relatively higher concentrations of eukaryotic-type chlorophylls that are produced from *Thalassionema*: 36.36% and *Pleurosigma*: 9.09%.,Dubai seawater has more bacterial-type chlorophylls although high percentages of *Thalassionema* and *Pleurosigma* are found (27.78% and 22.22%), which could be explained by the association of these species with bacteria that could perform the photosynthesis.

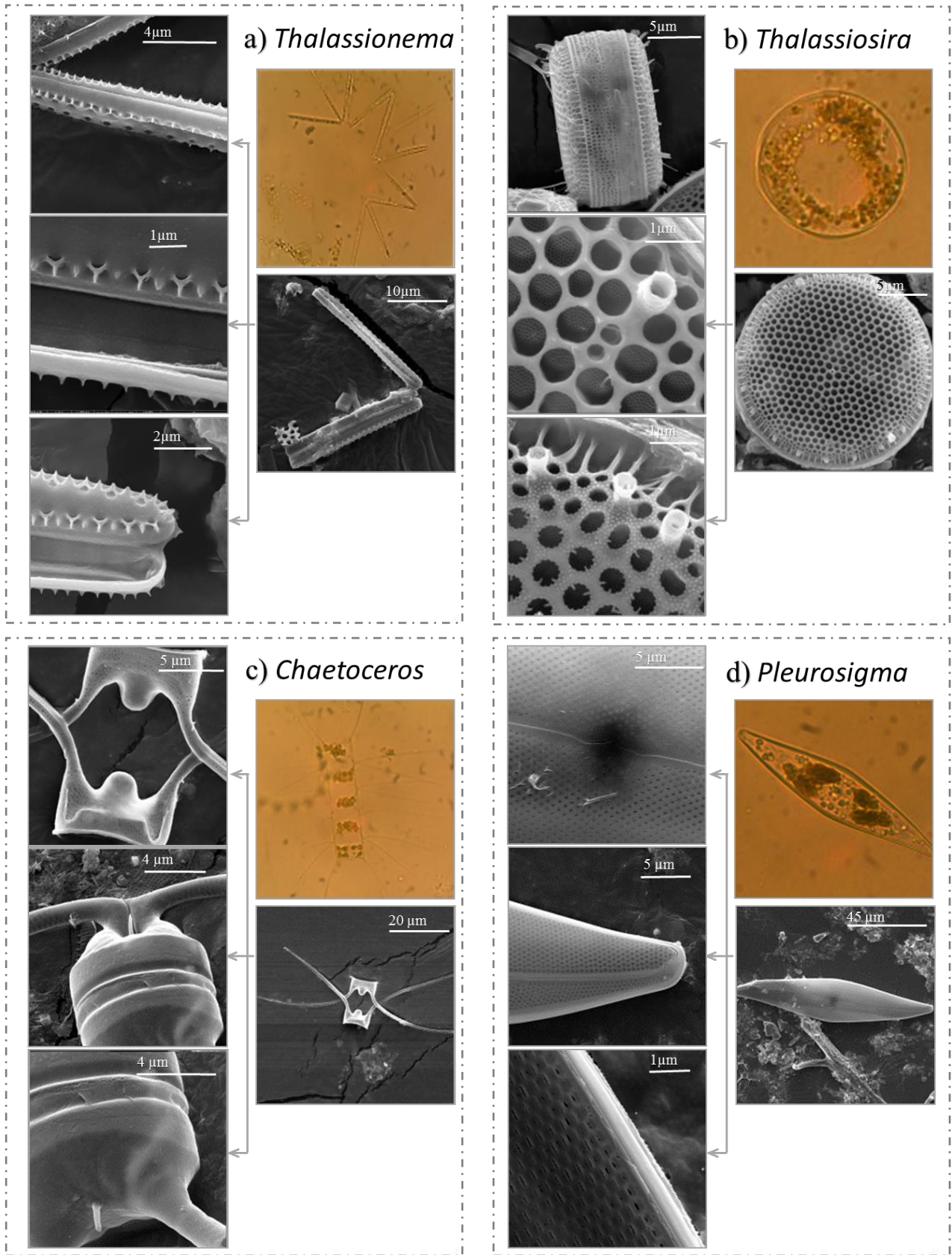

Figure 6. Light and electronic microscopic images of algal species causing some of blooms along the coasts of the United Arab Emirates.

Fucoxanthin and myristin were also detected in all the seawater samples collected. Fucoxanthin is an accessory pigment (carotenoid) that is usually found in the brown algae and diatoms (Peng et al., 2011) giving them a brown color. Our detection of relatively high levels of myristin, a saturated fat, indicated that this fat might be common among organisms inhabiting the UAE waters (Arabian Gulf). The plausibility of myristin-accumulation as a shared trait amongst Gulf organisms is increased when average summer temperatures approach 40 ºC. In this instance, the shorter, more highly saturated myristin could help protect against membrane instability that won't affect lipid bilayers with longer, polyunsaturated fatty acids. Pipercitine (an amide) was found at relatively high abundance in Sharjah seawater. In addition, an acetogenin, Sabadelin was detected at retention time of 2.7 min with [M+NH4]$^+$ ions m/z: 584.5. On the other hand, acetogenin was also found in the dinoflagellate blooms of UAQ seawater, which was dominated by *Noctulica* species.

Most of the seawater samples did not contain algal toxic compounds. However, the toxic compound Calicheamicinγ1 (Figure 7b) was detected at elevated concentrations in samples from Ajman. This toxic compound is commonly released from the bacteria *Micromonospora echinospora* (Griesbeck et al., 2012), which can be associated with the algal species in a symbiotic manner so their relationship with the algae is important.

Figure 7. HPLC-MS chromatograms: (a) versus m/z, and (b) versus retention time (RT) and structure of the Calicheamicinγ1.

3.4 The relation between the sea properties and algal blooms

The growth of the algal species is mainly supported by the chemical and physical properties of the waters where sea temperature and salinity can affect the metabolism of the algal cells and their nutrient uptake. Thus, the physical and chemical properties of the Arabian Gulf waters during bloom events and their relation with the algal abundances are described in the section as below:

*Temperature*

Sea temperature along the UAE coast shows high seasonal variability, with a peak of 36 °C in August and decreasing steadily to a minimum of around 11.9 °C in December (Al-Rashidi et al., 2009). This high seasonal variability significantly regulates the algal growth in the region. Figures 8-9 show the box plots of sea temperatures during the months in which algal blooms occur along the west and east coasts of the UAE. These figures show that algal blooms frequency increases at the temperature ranges between 23 and 24 °C in both east (Sea of Oman) and west coasts (Arabian Gulf). In addition, algal blooms can also occur at higher sea temperature ranges (>28 °C) exclusively in the west coast (Arabian Gulf), while they do not occur in the east coast at this temperature range. This is evident that there are several types of species that reside in the west coast, which can overgrow and adapt to the environment along this coast only.

To further investigate the influence of sea temperature on the algal growth in the west coast (Arabian Gulf), we have plotted the Chl-*a* versus the sea temperature of water samples collected along Abu Dhabi waters in 2010-2011 by EAD (Figure 10). This Figure highlights that there are two growth peaks of Chl-*a* occurring at 24 °C and 32 °C. This indicates that the algal growth is enhanced when sea temperature gets closer to the two 24 °C and 32 °C in the west coast (Arabian Gulf). However, in the east coast (Sea of Oman), the blooms usually grow in the colder waters below 28 °C. The optimum temperature ranges mentioned are found to be highly correlated with the photosynthetic rate that controls the algal growth (Vona et al., 2004). But, large deviations from the mentioned sea temperature ranges can reduce the algal growth rate due to limited enzyme activity, membrane fluidity, and nutrient uptake (Nejrup et al., 2013).

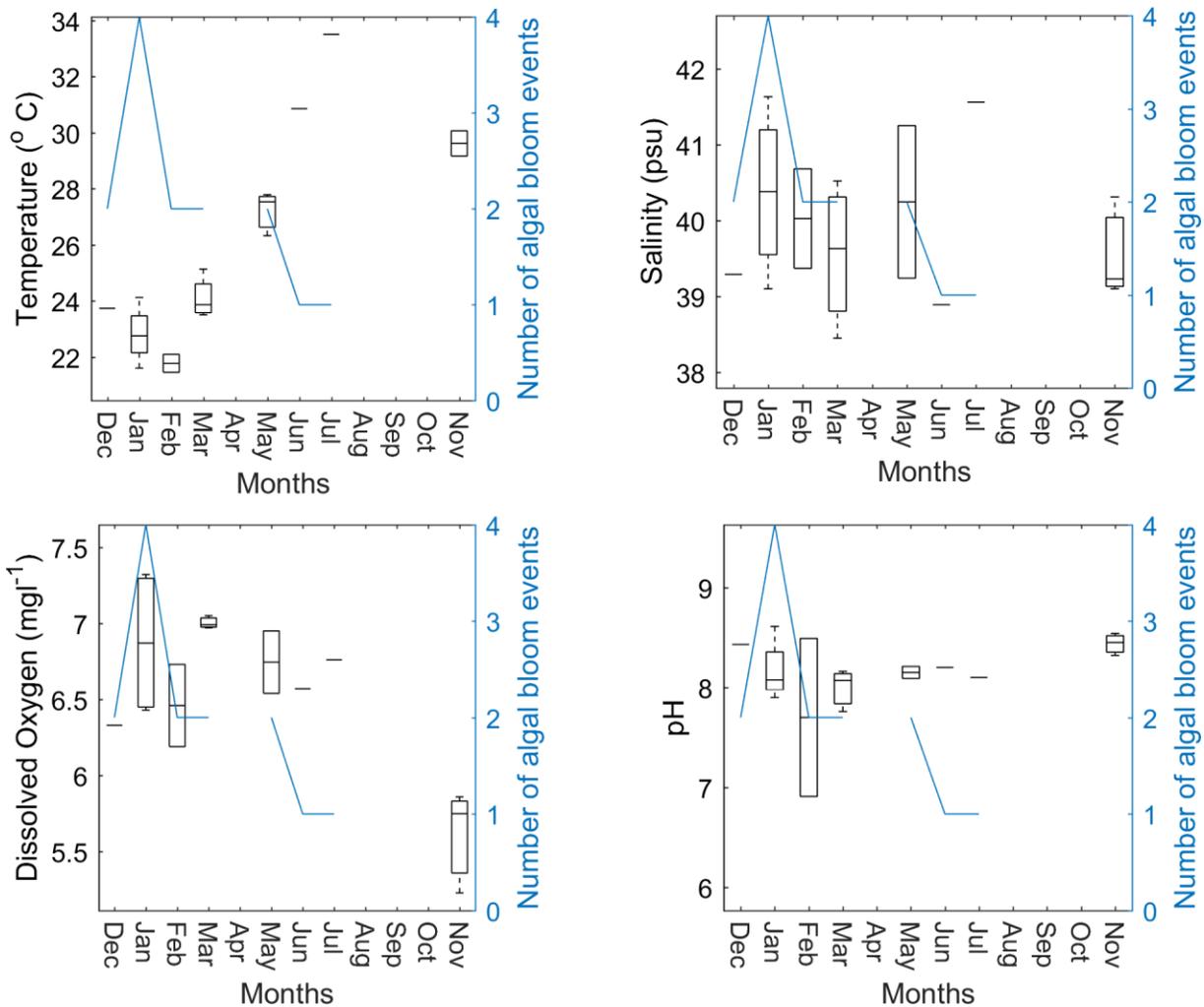

Figure 8. The frequency of algal bloom events, and the associated water properties (Temperature, Salinity, Dissolved Oxygen and pH) in the west coast (using MOCCAE data).

*Salinity*

In general, the waters of the east coast (Sea of Oman) are fresher (salinity 37 psu) than waters in the west coast (salinity ~40 psu) as shown in Figures 8-9. While the salinity shows a variability to some extent in the west coast (39-42 psu), it is mostly steady at 37.5 psu in the east coast due to its open boundary with Sea of Oman and Arabian Sea. This high variability of salinity in the east and west coasts could help in the growth of different kind of algal species. The highest number of blooms occur at salinity of (39-40 psu) in the west coast (Arabian Gulf), it occur at salinity of 37-37.5 psu in the east coast (Sea of Oman). Blooms occur at these salinity ranges because they are within the optimum salinity range of 24-40 psu where the maximum yield of algal species could be achieved (Gireesh et al., 2008; Sigaud and Aidar, 1993). This is due to the dependency of the algal growth rate, maximum yield and Chl-*a* content on the algal species and the salinity (Sigaud and Aidar, 1993). The decrease in salinity can inhibit cell division and increase the amount of Chl-*a* (Sigaud and Aidar, 1993). On the other hand the increase in salinity causes an increase in carotenoid content in some of the algal species (Borowitzka et al., 1990; Rao Ranga et al., 2007).

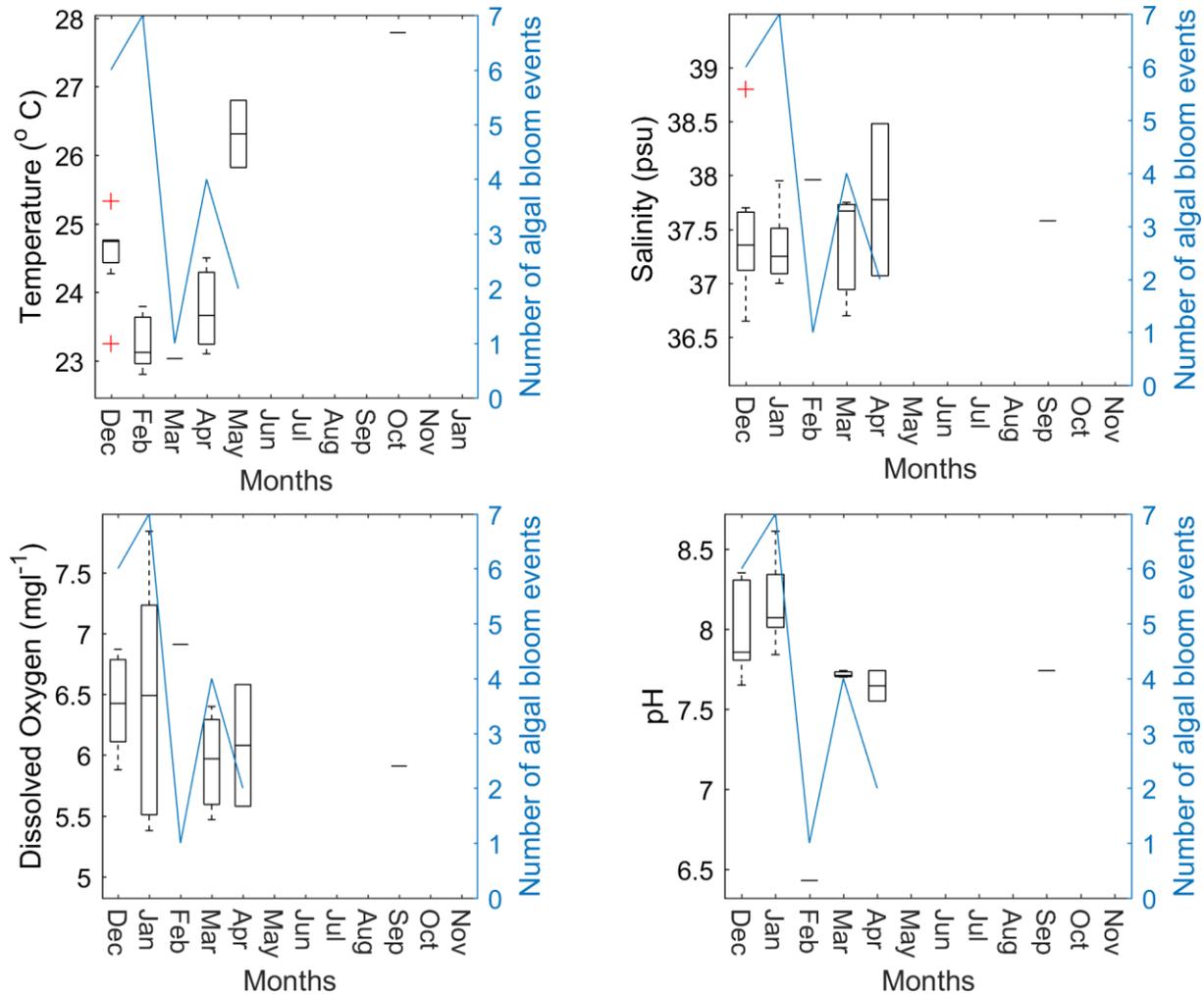

Figure 9. The frequency of algal bloom events, and the associated water properties (Temperature, Salinity, Dissolved Oxygen and pH) in the east coast (using MOCCAE data).

*Dissolved Oxygen*

In the west waters of the UAE along the Arabian Gulf, high dissolved oxygen level is found (reaching up to 7.5 mg L$^{-1}$), but in the east coast waters (Sea of Oman) lower levels of dissolved oxygen is observed and mostly less than 6.5 mg L$^{-1}$ during the whole year. As shown in Figures 8-9, the blooms occur commonly at the oxygen levels of 6.5-7 mg L$^{-1}$ and 6-6.5 mg L$^{-1}$ in the west and east waters, respectively. However, it is very interesting to see that the initial annual bloom events occurring in November and January in the west and east coast respectively coincide with the minimum levels of oxygen in their respective waters. This does mean that after long summer period where no blooms have occurred, the blooms start their biological year with consuming the available oxygen for their growth. After this period, the blooms start contributing to producing the oxygen again and causing an increase in the level of oxygen in the water in the advanced blooming period.

To further investigate the relation between the algal blooms and oxygen levels in the water, we have plotted the concentration of Chl-*a* and dissolved oxygen measured along the west coast of UAE (Abu Dhabi) during the years 2010-2011 (Figure 10). We found that, low Chl-*a* (less than 5 mg m$^{-3}$) is always concurring at oxygen levels between 4 to 6 mg.L$^{-1}$. However, there are two interesting conditions observed for the high Chl-*a* concentration (higher than 5 mg.m$^{-3}$). Clustering methods have been used to differentiate between these cases. The first case found is that when the dissolved oxygen level is higher than 6 mg.L$^{-1}$, the Chl-*a* is always higher than 5 mg.m$^{-3}$ approaching 15 mg.m$^{-3}$. This condition is matching the advanced blooming period mentioned above when algal species respire and increase the production of oxygen in the water. However, the other condition is that when the dissolved oxygen level is lower than 4 mg.L$^{-1}$, the Chl-*a* is also found to be higher than 5 mg.m$^{-3}$ and exceeding 20 mg.m$^{-3}$. This condition happens at the initial blooming conditions when algal blooms try to survive and increase in its growth rate. Based on these observations, we can state that if the oxygen level is below 4 mg.L$^{-1}$ and above 6 mg.L$^{-1}$, it is an indication for algal blooms in the west coast of UAE. However, in the case of oxygen level falling between 4 and 6 mg.L$^{-1}$, Chl-*a* cannot be predicted and it is a transition period of both less and more algal blooming.

*pH*

The pH is around 8 at the west algal blooms, and it varies between 6.5 and 8.5 in the east algal bloom events. However, while there is no significant changes of algal bloom frequencies with respect to the pH level in the west coast, it is very interesting to see that frequent algal blooms occur at the pH levels around 8 in the east coast as shown in Figures 8-9. Therefore, it is pertinent to note that 8 is the optimum pH level for the algal blooms to occur in the west and east coasts of the UAE.

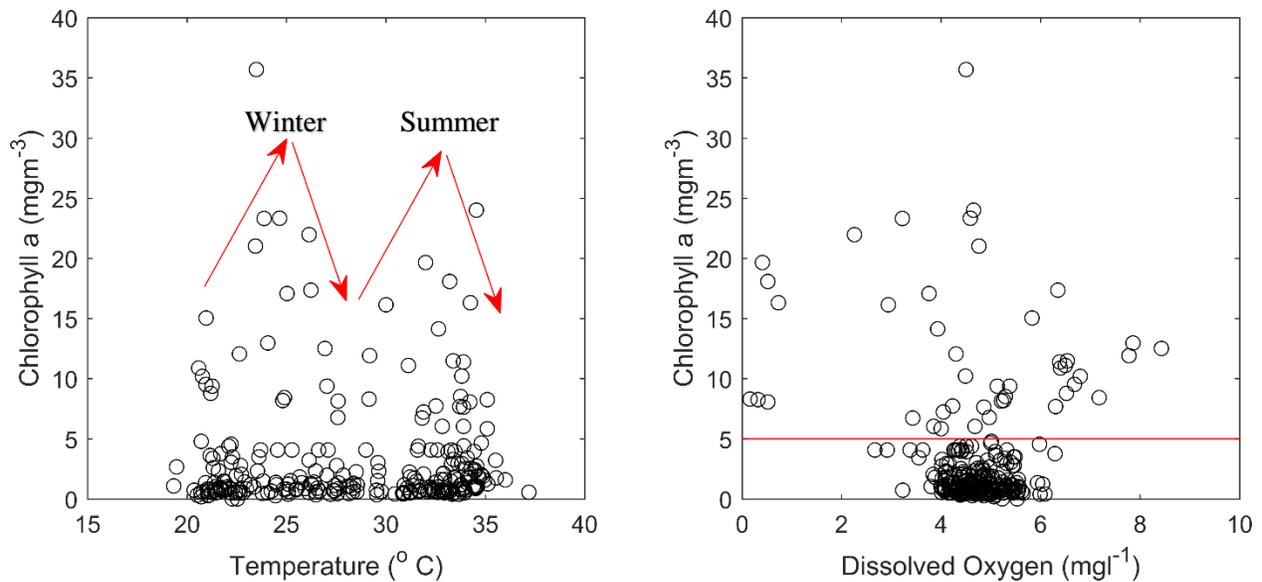

Figure 10. Scatter plots of Chlorophyll a versus temperature and Dissolved oxygen of Abu Dhabi waters (using EAD data).

## 4. Conclusions

Our analysis shows that events occur most frequently in the winter and spring season in the shallow waters of Arabian Gulf and deep waters of Sea of Oman. Whereas the lowest number of algal blooms occur in the summer. There is a general decreasing trend of the algal bloom events from 2010 to 2018 in the Arabian Gulf whereas there is an increasing trend of algal blooms events in the Sea of Oman over the same period. We have noticed that in this region have a distinct feature that the initial blooming happens in November-December and December-January in the Arabian Gulf and Sea of Oman, respectively. We found that Algal bloom events are related to water properties such hat they grow at specific water conditions (pH level and temperatures). For instance, the algal blooms frequently occur in water with pH at 8 in both Arabian Gulf and Sea of Oman. This means that the algal blooms grow at the same pH conditions in both the shallow and deep waters. In addition, algal growth is enhanced when sea temperature gets closer to 24 ºC and 32 ºC in the shallow waters of Arabian Gulf. However, in the deep waters of Sea of Oman, the blooms usually grow in the colder waters below 28 ºC. We have also found that, while the highest number of blooms occur at salinity of 39-40 psu in the shallow Arabian Gulf, it occur at salinity of 37-37.5 psu in the fresher deeper waters of Oman Sea. Future research should seek to identify the drivers behind the observed spatio-temporal trends in algal bloom event occurrence in this region, and aim to further explain the significant correlations between the nutrients levels and algal blooms.